\documentclass[conference]{IEEEtran}
\usepackage{graphicx}
\usepackage{amsmath}
\usepackage[a4paper,bindingoffset=0.2in,%
            left=0.65in,right=0.65in,top=0.75in,bottom=0.95in,%
            footskip=.25in]{geometry}
\usepackage{lipsum}
\usepackage{amsthm}
\usepackage[strings]{underscore}
\usepackage{amsfonts}
\usepackage{caption}
\usepackage{siunitx}
\usepackage{amssymb}
\usepackage{subcaption}
\usepackage{placeins}
\usepackage{lettrine}
\usepackage{multicol, blindtext}
\usepackage{bbm}
\usepackage{cite}
\usepackage{color}
\usepackage{algorithm}
\usepackage{algpseudocode}
\usepackage{authblk}
\newtheorem{remark}{Remark}
\makeatletter

\newcommand{\Rmnum}[1]{\expandafter\@slowromancap\romannumeral #1@}
\newenvironment{psmallmatrix}
  {\left[\begin{smallmatrix}}
  {\end{smallmatrix}\right]}
\newtheorem{theorem}{Theorem}

\newtheorem{proposition}{Proposition}

\newtheorem{lemma}{Lemma}
\makeatletter
\def\BState{\State\hskip-\ALG@thistlm}
\makeatother
\ifCLASSINFOpdf
\else
\fi

\begin{document}
%
\title{Minimizing The Age of Information in a CSMA Environment}
\author[*]{Ali Maatouk}
\author[*]{Mohamad Assaad}
\author[$\dagger$]{Anthony Ephremides}
\affil[*]{TCL Chair on 5G, Laboratoire des Signaux et Syst\`emes, CentraleSup\'elec, Gif-sur-Yvette, France }
\affil[$\dagger$]{ECE Dept., University of Maryland, College Park, MD 20742}
\maketitle

\begin{abstract}
In this paper, we investigate a network of $N$ interfering links contending for the channel to send their data by employing the well-known Carrier Sense Multiple Access (CSMA) scheme. By leveraging the notion of stochastic hybrid systems, we find a closed form of the total average age of the network in this setting. Armed with this expression, we formulate the optimization problem of minimizing the total average age of the network by calibrating the back-off time of each link. By analyzing its structure, the optimization problem is then converted to an equivalent convex problem that can be solved efficiently to find the optimal back-off time of each link. Insights on the interaction between the links is provided and numerical implementations of our optimized CSMA scheme in an IEEE 802.11 environment is presented to highlight its performance. We also show that, although optimized, the standard CSMA scheme still lacks behind other distributed schemes in terms of average age in some special cases. These results suggest the necessity to find new distributed schemes to further minimize the average age of any general network.
\end{abstract}


%
\IEEEpeerreviewmaketitle

\section{Introduction}
\lettrine{S}{}ince its introduction in \cite{6195689}, the Age of Information (\textbf{AoI}) has been capturing wide attention and is considered of ample importance in communication networks. With new technologies emerging and the abondance of cheap hardware, constant connectivity and monitoring has been the norm in the modern technology applications. In these applications, updates about a specific metric of interest is sent from an information source to a monitor (e.g. temperature of a room\cite{5597912}, vehicle's position and velocity \cite{5307471} etc.). More specifically, this source produces time-stamped status updates that are sent through the network towards the intended receiver. The ultimate goal becomes for the monitor to have the freshest knowledge about the metric of interest. However, the investigation of the AoI, even in the simplest scenarios, was shown to be challenging and far from straightforward.

Since the seminal work in \cite{6195689}, the AoI has been gaining the attention of researchers and a large amount of papers have been published on this subject. The seminal work in \cite{6195689} investigated the average age in First-Come-First-Served (\textbf{FCFS}) disciplines: M/M/1, M/D/1 and D/M/1. In \cite{6875100}, packets management was shown to further minimize the average age of the stream of interest and a related metric, the \emph{average peak age}, was introduced. The average age was also investigated in sampling problems. For instance, it was shown in \cite{8000687} that in a $``$generate-at-will$"$ model, a zero-wait policy is not always optimal in minimizing the average age. The AoI has also gained wide attention for energy harvesting sources (e.g. \cite{2018arXiv180202129A}). Recently, multi-hop scenarios were investigated in \cite{8262777,8445981,8006593} where in the latter, it was shown that the Last-Come-First-Served (\textbf{LCFS}) discipline at relaying nodes minimizes the average age of the considered stream. There have also been an increased interest in the study of the AoI in prioritized queuing \cite{2018arXiv180805738Z,2018arXiv180104067N,2018arXiv180511720M}. For instance, in \cite{2018arXiv180805738Z}, the AoI in a multicast network where receivers are divided into two groups: priority and non-priority group was investigated. Lately, scheduling in the aim of minimizing the average age of the network has been considered (e.g. \cite{8006590}). However, the focus of the literature in this regards lays mainly on centralized scheduling schemes and theoretical work on distributed scheduling schemes is limited. With the AoI being of broad interest in sensors-type applications where distributed schemes are needed (e.g. environmental monitoring and vehicular networks \cite{5597912}\cite{5307471}), minimizing the total average age in a distributed environment is therefore of paramount importance. In \cite{2018arXiv180306469T}, the authors studied a distributed scheme where devices, that always have fresh information at their disposal, access the channel with a certain probabiblity. The access probability of each link is then optimized in the aim of minimizing the total average age of the network. In \cite{2018arXiv180103975J}, it was shown that a Round Robin multiple access scheme is asymptotically optimal (when devices' numbers tends to infinity) in minimizing the AoI. However, the work in \cite{2018arXiv180103975J} assumed that the transmission of \emph{all} links is the \emph{same} \emph{deterministic} quantity. As links usually exhibit different random channel conditions, and as packets may arrive stochastically to the devices, in our paper we take a general scenario where: 1) devices have stochastic packets arrivals 2) the transmission time of each link is supposed to be \emph{random} and with \emph{different} average transmission time for each link. Moreover, our work focuses on Carrier Sense Multiple Access (\textbf{CSMA}), which is regarded as one of the most practical distributed Multiple Access Protocol (\textbf{MAC}) schemes in wireless networks. 

CSMA is a class of simple and distributed multiple access protocol that is seen as one of the most popular distributed MAC schemes in wireless networks (e.g. CSMA is the basic medium access
algorithm in IEEE 802.11). In this class of schemes, a transmitter attempts to determine whether another transmission is in progress using a carrier-sense mechanism before initiating a transmission. There exist a vast amount of research results on CSMA in regards of its theoretical analysis and applications. For example, throughput optimal CSMA schemes have been extensively investigated in the literature (we refer the readers to \cite{6406138} for a survey). Propositions of CSMA schemes where the battery limitation of devices is taken into account has also been massively studied in previous works \cite{1632658}. For example, the authors in \cite{DBLP:journals/corr/abs-1712-03063} proposed a throughput optimal CSMA scheme where the energy aspect of the transmission is taken into account\color{black}. With the AoI metric being relatively new, there have only been little work on CSMA in the AoI literature. More specfically, CSMA has been investigated in the framework of vehicular networks in \cite{5984917}. In the aforementioned reference, the authors studied the average AoI of a vehicular network environment numerically. It was shown that the
information age is minimized at an optimal operating point (optimal average back-off time) that
lies between the two extremes of maximum throughput and minimum
delay. However, closed form results on that optimal point was not provided. Based on the preceding, a concrete theoretical analysis to find the optimal back-off time remains an open question.
This motivated our work where we first leverage the notion of Stochastic Hybrid Systems (\textbf{SHS}) to model a CSMA environment in which $N$ interfering links contend for the channel. Armed with these tools, we provide a closed form of the total average age of the network in function of the arrival rate, the average transmission time and the average back-off time of each link. Afterwards, we formulate our optimization problem in the aim of minimizing the total average age of the network. This problem is then shown to have an equivalent convex formulation that can be solved efficiently to find the optimal back-off time of each link. Theoretical insights on the interaction between links is then provided. After that, our proposed solution is implemented in a realistic IEEE 802.11 setting and its performance is highlighted. To the knowledge of the authors, this is the first work that provides theoretical results on the optimal back-off time of links in a CSMA setting with the aim of minimizing the total average AoI of the network. Furthermore, we take a special scenario and compare the optimized CSMA scheme to other distributed schemes in terms of average AoI. Interestingly, we show that the optimized CSMA scheme still lacks behind in terms of average AoI with respect to the other schemes. This is in contrast to the maximization of throughput literature where standard CSMA schemes were shown to achieve the maximum achievable throughput of the network when optimized \cite{6406138}. Therefore, these results suggest the necessity of the foundation of new distributed schemes to further minimize the average AoI in any general network.

The paper is organized as follows: Section \Rmnum{2} presents the system model. Section \Rmnum{3} provides the theoretical findings on the average age and the optimization problem in question. Section \Rmnum{4} lays out the numerical results that corroborate the theoretical findings while Section \Rmnum{5} concludes the paper.

\section{System Model}
We consider in our paper $N$ links (transmitter-receiver pairs) sharing the medium of transmission. The transmitter side of each link send status updates to its corresponding monitor. However, due to interference, only one link can be active at each time instant. We suppose that the packets arrival at each transmitter $k$ is exponentially distributed with a rate of $\lambda_k$. The transmission time of the packets is assumed to be exponentially distributed with an average channel holding time of $\frac{1}{H_k}$. We forgo queuing in this network; each transmitter keeps at most one packet in its system. Upon a new arrival of a packet to transmitter $k$, the packet of transmitter $k$ that is currently available (or being served) is preempted and discarded\footnote{To make the model more tractable, it is assumed that a transmitter that captures the medium, sends a ``fake" update (i.e. a packet with the same time stamp as the previously transmitted packet) if its buffer is empty \cite{DBLP:journals/corr/YatesK16}}. This setting is motivated by the fact that a preemptive M/M/1/1 scenario was shown to minimize the average age in the case of exponential transmission time \cite{8006593}. Moreover, this model is useful in realistic scenarios (e.g. congested wireless network) as the service time would be dominated by the MAC access delay \cite{2018arXiv180307993Y}. 

This paper focuses on a network where links employ CSMA to access the channel. In this class of schemes, a transmitter in the network listens to the medium before proceeding to the transmission. More specifically, a transmitter waits for a certain duration of time before transmitting, called the \emph{back-off} time. While waiting, it keeps sensing the environment to spot any conflicting transmission. If an interfering transmission is spotted, the transmitter stops immediately its back-off timer and waits for the medium to be free to resume it. We suppose that the chosen back-off timer of link $k$ is exponentially distributed with an average of $\frac{1}{R_k}$. This assumption is motivated by the fact that exponentially distributed back-off timers were shown to be throughput optimal (i.e. can achieve the capacity region of the network \cite{DBLP:journals/corr/abs-1712-03063}\cite{5340575}). In the following, we adopt the standard idealized CSMA assumptions \cite{5340575}:
\begin{itemize}
\item The problem of \emph{hidden nodes} does not exist
\item Sensing is considered instantaneous, there is no sensing delay
\end{itemize}
The first condition is satisfied in realistic scenarios if the range of carrier-sensing is large enough. On the other hand, the second condition is violated in practical systems due to the time needed for a receiver to detect the radio signals. However, this condition simplifies the mathematical model, making it tractable, which can be a starting point before investigating the cases where this condition is violated. In fact, with these idealized CSMA settings and the continuous nature of the back-off timer, collisions become mathematically impossible. This leads us to capture the essence of the scheduling problem in question without being concerned about the contention resolution problem. We'll discuss in Section \Rmnum{3}-C how the present work and analysis can be easily extended to the case where contention resolution is involved and we will provide implementation considerations in IEEE 802.11 networks.

In this scenario, the instantaneous age of information at the receiver (monitor) of link $k$ at time instant $t$ is defined as:
\begin{equation}
\Delta_k(t)=t-U_k(t)
\end{equation}
where $U_k(t)$ is the time stamp of the last successfully received packet by the receiver side of link $k$. Clearly the evolution of the age will depend on the arrival process of each link, along with the time spent to capture the medium (back-off mechanism) and the transmission time. The ultimate goal therefore consists of minimizing the total average age of the network that is defined as:
\begin{equation}
\overline{\Delta}=\sum_{k=1}^{N}\overline{\Delta}_k=\sum_{k=1}^{N}\lim_{\tau\to\infty}\frac{1}{\tau}\int_{0}^{\tau}\Delta_k(t)dt
\end{equation}
\section{Theoretical Analysis}
\subsection{Preliminaries on SHS}
The standard approach to calculate the average age throughout the literature is a graphical one which revolves around decomposing the area below the curve of the instantaneous age to multiple trapezoids and calculating each of their area \cite{6195689}. However, this approach can be quite challenging in queuing systems where packets can be dropped (i.e lossy systems) \cite{2018arXiv180307993Y}. We therefore adopt the SHS approach to model and calculate the total average of the system in question \cite{DBLP:journals/corr/YatesK16}. The SHS approach consists of modeling the system through the states $(q(t),\boldsymbol{x}(t))$ where $q(t)\in\mathbb{Q}$ is a discrete process that captures the evolution of the system in question while $\boldsymbol{x}(t)\in\mathbb{R}^{2N}$ is a continuous process that captures the evolution of the age process of each link $k$ at the monitor along with the packet present in the system at each link $k$. More specifically:
\begin{equation}
\boldsymbol{x}(t)=[x_0(t),x_1(t),\ldots,x_{2N-2}(t),x_{2N-1}(t)]
\end{equation}
where:
\begin{equation*}
\begin{cases}
x_{2k}(t)& \text{Age of link}\: $k+1$\: \text{at the monitor} \: 0\leq k\leq N-1\\
x_{2k+1}(t)& \text{Age of the packet at link} \: $k+1$ \:\:\: 0\leq k\leq N-1\\
\end{cases}
\label{Stationary}
\end{equation*}
For completeness, we \emph{briefly} describe the idea of SHS and we refer the readers to \cite{DBLP:journals/corr/YatesK16} for a more in-depth discussion.

The discrete process $q(t)$ is a Markov process that can be represented graphically by a Markov chain $(\mathbb{Q},\mathbb{L})$ in which each state $q\in\mathbb{Q}$ is a vertex in the chain and each transition\footnote{It is worth mentioning that unlike a typical continuous-time Markov chain, the chain $(\mathbb{Q},\mathbb{L})$ may include self-transitions where a reset of the continuous process $\boldsymbol{x}$ takes place but the discrete state remains the same.
} $l\in\mathbb{L}$ is a directed edge $(q_l,q'_{l})$ with a transition rate $\lambda^{(l)}\delta_{q_l,q(t)}$ where the Kronecker delta function assures that this transition $l$ can only take place when the discrete process $q(t)$ is equal to $q_l$. For each state $q$, we define the incoming and outgoing transitions sets respectively as:
\begin{equation}
\mathbb{L}'_q=\{l\in\mathbb{L}: q'_{l}=q\} \quad \mathbb{L}_q=\{l\in\mathbb{L}: q_{l}=q\}
\end{equation}
The interest in SHS comes from the fact that the discrete process transitions will result in a reset mapping in the continuous process. More specifically, when a transition $l$ takes place, the discrete process jumps to another state $q'_{l}$ and a discontinuous jump in the continuous process $\boldsymbol{x'}=\boldsymbol{x}\boldsymbol{A_l}$ is induced. The matrix $\boldsymbol{A_l}\in\mathbb{R}^{2N}\times\mathbb{R}^{2N}$ is referred to as the transition reset maps. Finally, at each state $q\in\mathbb{Q}$, the continuous process evolves through the following differential equation $\dot{\boldsymbol{x}}=\boldsymbol{b}_q$ where $b^k_q$ is a binary element that is equal to $1$ if the age process $x_k$ increases at a unit rate when the system is in state $q$ and is equal to $0$ if it keeps the same value.

To calculate the average age of the system through SHS, the following quantities for each state $q\in\mathbb{Q}$ need to be defined:
\begin{equation}
\pi_{q}(t)=\mathbb{E}[\delta_{q,q(t)}]=P(q(t)=q)
\end{equation}
\begin{equation}
\boldsymbol{v}_{q}(t)=[v_{q0}(t),\ldots,v_{q2N}(t)]=\mathbb{E}[\boldsymbol{x}(t)\delta_{q,q(t)}]
\end{equation} 
where $\pi_{q}(t)$ is the Markov chain's state probability and $\boldsymbol{v}_{q}(t)$ denotes the correlation between the age process and the discrete state of the system $q$. One of the key assumption is that the Markov chain $q(t)$ is ergodic and therefore we can define the steady state probability vector $\overline{\boldsymbol{\pi}}$ as the solution to the following equations:
\begin{equation}
\overline{\pi}_{q}(\sum_{l\in\mathbb{L}_q}\lambda^{(l)})=\sum_{l\in\mathbb{L}'_q}\lambda^{(l)}\overline{\pi}_{q_l} \quad q\in\mathbb{Q}
\end{equation}
\begin{equation}
\sum_{q\in\mathbb{Q}}\overline{\pi}_{q}=1
\end{equation}
As it has been shown in \cite{DBLP:journals/corr/YatesK16}, the correlation vector $\boldsymbol{v}_{q}(t)$ converges in this case to a limit $\overline{\boldsymbol{v}}_{q}$ such that:
\begin{equation}
\overline{\boldsymbol{v}}_{q}(\sum_{l\in\mathbb{L}_q}\lambda^{(l)})=\boldsymbol{b}_q\overline{\pi}_{q}+\sum_{l\in\mathbb{L}'_q}\lambda^{(l)}\overline{\boldsymbol{v}}_{q_l}\boldsymbol{A_l} \quad q\in\mathbb{Q}
\label{solutionv}
\end{equation}
Based on this, we can deduce that $\mathbb{E}[x_{2k}]=\lim\limits_{t \to +\infty} \mathbb{E}[x_{2k}(t)]=\lim\limits_{t \to +\infty}\sum\limits_{q\in\mathbb{Q}}\mathbb{E}[x_{2k}(t)\delta_{q,q(t)}]=\sum\limits_{q\in\mathbb{Q}}\overline{v}_{q2k}\:\:\forall k\in\{0,\ldots,N-1\}$.\\
Based on the aforementioned results from \cite{DBLP:journals/corr/YatesK16} and as the ultimate goal is to calculate the average age at the monitor of each link $k$, we present the following theorem that summarizes all what have been previously stated:
\begin{theorem}
When the markov chain $q(t)$ is ergodic and admits $\boldsymbol{\overline{\pi}}$ as stationary distribution, if we can find a solution for eq. (\ref{solutionv}), then the average age at the monitor of link $k \:\:\forall k\in\{1,\ldots,N\}$, is:
\begin{equation}
\overline{\Delta}_k=\sum_{q\in\mathbb{Q}}\overline{v}_{q2(k-1)}
\end{equation}
\label{theomhem}
\end{theorem}
\subsection{Average age calculation}
In order to simplify the average age calculations, we forgo studying the age process of all links simultaneously. Instead, we examine a link of interest $i$ and calculate its average age by considering the network from its perspective. In this case, we define the discrete states set $\mathbb{Q}=\{0,1,2,\ldots,N\}$ where $q(t)=k$ if link $k$ has captured the channel and started transmission at time $t$ while $q(t)=0$ if the channel is idle at time $t$. The continuous-time state process is defined as $\boldsymbol{x}(t)=[x_{0}(t),x_{1}(t)]$ where $x_{0}(t)$ is the age of the link of interest $i$ at the monitor at time $t$ and $x_{1}(t)$ is the age of the packet in the system of link $i$ at time $t$. Our goal becomes to apply Theorem \ref{theomhem} to conclude the vectors $\overline{\boldsymbol{v}}_{q}=[\overline{v}_{q0},\overline{v}_{q1}]\:\: \forall q\in\mathbb{Q}$ that will allow us to calculate the average age of the link of interest $i$.

To proceed towards our goal, we summarize the transitions between the discrete states and the reset maps they induce on the age process $\boldsymbol{x}(t)$ in this table:
\begin{center}
\begin{tabular}{cccccc}
$l$ & $q_l\rightarrow q'_{l}$ & $\lambda^{(l)}$ & $\boldsymbol{xA_l}$ & $\boldsymbol{A_l}$  &  $\boldsymbol{v}_{q_l}\boldsymbol{A_l}$ \\
\hline
 $1$ & $0\rightarrow1$  & $R_1$ & $[x_0,x_1]$ & $\begin{psmallmatrix}
    1 & 0  \\
    0& 1
\end{psmallmatrix}$  &  $[v_{00},v_{01}]$\\
 $2$ & $0\rightarrow2$ & $R_2$  & $[x_0,x_1]$ & $\begin{psmallmatrix}
    1 & 0  \\
    0& 1
\end{psmallmatrix}$  &  $[v_{00},v_{01}]$\\
    & $\vdots$ & $\vdots$ & $\vdots$ & $\vdots$  &  $\vdots$\\
 $N$ &  $0\rightarrow N$  & $R_N$ & $[x_0,x_1]$ & $\begin{psmallmatrix}
    1 & 0  \\
    0& 1
\end{psmallmatrix}$  &  $[v_{00},v_{01}]$\\
  $N+1$ &  $1\rightarrow0$  & $H_1$  & $[x_0,x_1]$ & $\begin{psmallmatrix}
    1 & 0  \\
    0& 1
\end{psmallmatrix}$   &  $[v_{10},v_{11}]$ \\
  $N+2$ &  $2\rightarrow0$  & $H_2$  & $[x_0,x_1]$ & $\begin{psmallmatrix}
    1 & 0  \\
    0& 1
\end{psmallmatrix}$   &  $[v_{20},v_{21}]$ \\
      & $\vdots$ & $\vdots$ & $\vdots$ & $\vdots$  &  $\vdots$\\
      $N+i$ &  $i\rightarrow0$  & $H_i$ & $[x_1,x_1]$ & $\begin{psmallmatrix}
    0 & 0  \\
    1& 1
\end{psmallmatrix}$   &  $[v_{i1},v_{i1}]$ \\
      & $\vdots$ & $\vdots$ & $\vdots$ & $\vdots$  &  $\vdots$\\
      $2N$ &  $N\rightarrow0$  & $H_N$ & $[x_0,x_1]$ & $\begin{psmallmatrix}
    1 & 0  \\
    0& 1
\end{psmallmatrix}$  &  $[v_{N0},v_{N1}]$ \\
      $2N+1$ &  $0\rightarrow0$  & $\lambda_i$  & $[x_0,0]$ & $\begin{psmallmatrix}
    1 & 0  \\
    0& 0
\end{psmallmatrix}$   &  $[v_{00},0]$ \\
      $2N+2$  &  $1\rightarrow1$  & $\lambda_i$ & $[x_0,0]$ & $\begin{psmallmatrix}
    1 & 0  \\
    0& 0
\end{psmallmatrix}$   & $[v_{10},0]$ \\
            & $\vdots$ & $\vdots$ & $\vdots$ & $\vdots$  &  $\vdots$\\
     $3N+1$ & $N\rightarrow N$ & $\lambda_i$ & $[x_0,0]$ & $\begin{psmallmatrix}
    1 & 0  \\
    0& 0
\end{psmallmatrix}$   &  $[v_{N0},0]$ \\
\end{tabular}
 \captionof{table}{Stochastic Hybrid System Description}
 \label{trans}
\end{center}
In the following, we elaborate on the transitions reported in Table \ref{trans}:
\begin{enumerate}
\item The first set of transitions spanning from $l=1$ till $l=N$ corresponds to the case where link $k$ captures the channel. The rate of each of these transitions $l=k$ is $R_k$. Capturing the channel have no effect on the age process at the monitor of the link of interest $i$ nor on the packet in the system and the age process vector therefore stays the same without any reset $\boldsymbol{x'}=\boldsymbol{xA_l}=\boldsymbol{x}$.
\item The second set of transitions spanning from $l=N+1$ till $l=2N$ corresponds to the case where link $k$ releases the channel upon a successful transmission. As it can be seen, a successful transmission of any packet belonging to links $k\neq i$ will not result in any reset of the age process $\boldsymbol{x}$. On the other hand, for the transition $l=N+i$, the age process at the monitor of link $i$ resets to the age of the packet that was delivered $x_1$. As it has been previously explained in the paper, to avoid tracking the buffer status of link $i$, we suppose a ``fake" update is generated with the same age of the previously transmitted packet $x_1$ and therefore $x'_1=x_1$.
\item The final set of transitions spanning from $l=2N+1$ till $l=3N+1$ corresponds to a new packet arrival for link $i$. This new packet will replace the packet that is already in the system that is currently being served or waiting to be served. In other words, the new packet arrival will keep the same age at the monitor $x_0$ but will reset the age of the system's packet at link $i$ to $0$.
\end{enumerate}
\begin{figure}[!ht]
\centering
\includegraphics[width=.8\linewidth]{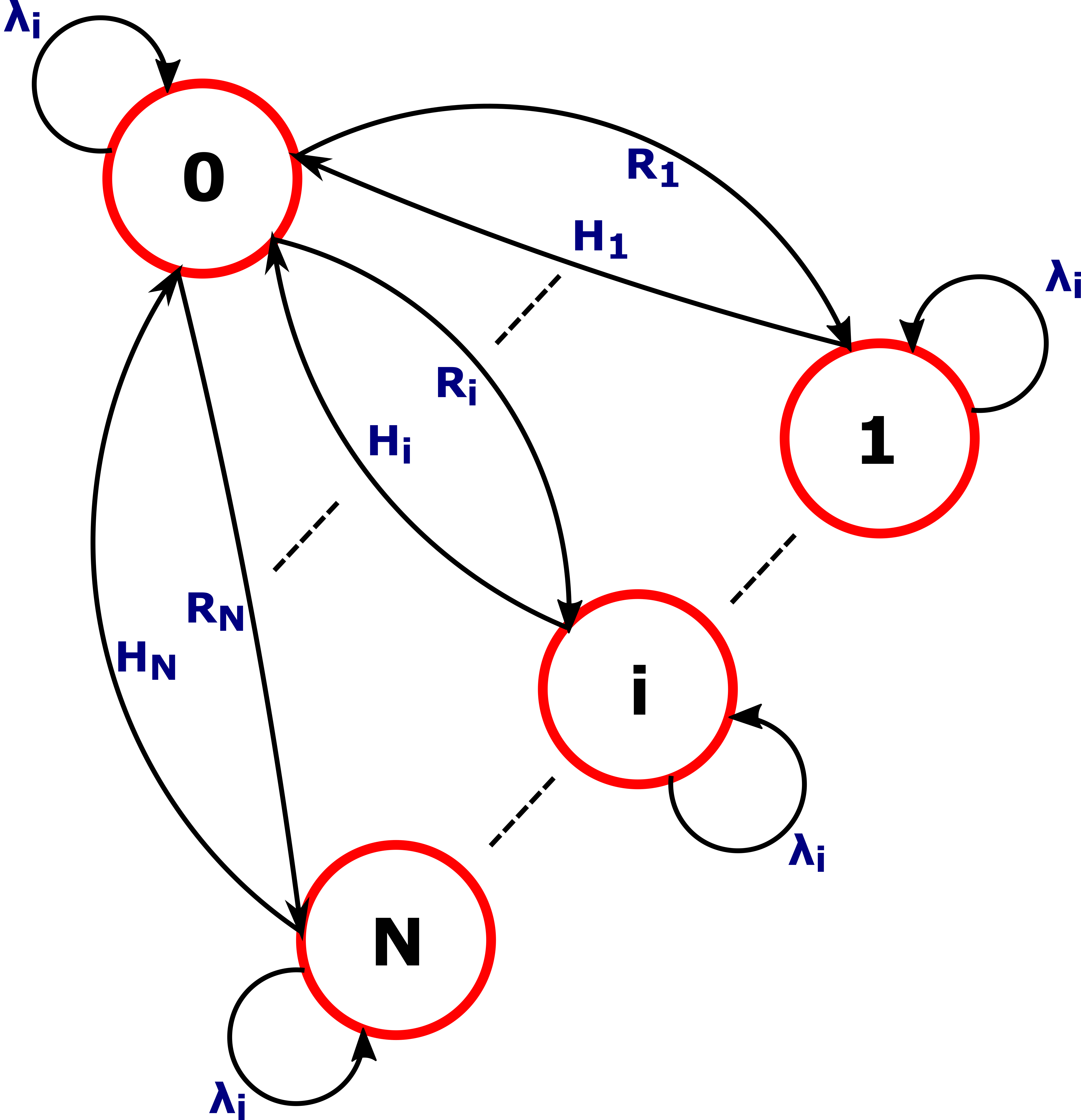}
\setlength{\belowcaptionskip}{-10pt}
\caption{Illustration of the stochastic hybrid systems Markov chain}
\label{effect}
\end{figure}
As for the differential equations governing the evolution of the age process in each discrete state, we have that in each state $q\in\mathbb{Q}$, both $x_{0}(t)$ and $x_{1}(t)$ increase at a unit rate:
\begin{equation}
\boldsymbol{b}_q=[1\:\:1] \quad\forall q\in \mathbb{Q}
\label{bstuff}
\end{equation}
To proceed with applying Theorem \ref{theomhem}, we first have to find the stationary distribution of the Markov Chain that models the transitions reported in Table \ref{trans}. For this purpose, we provide the following proposition:
\begin{proposition} The continuous time Markov chain is irreducible, time-reversible and admits $\overline{\pi}(k;\boldsymbol{R})$ as stationary distribution for any feasible state $0\leq k\leq N$ where:
\begin{equation}
\begin{cases}
\overline{\pi}(0;\boldsymbol{R})=\frac{1}{C(\boldsymbol{R})}\\
\overline{\pi}(k;\boldsymbol{R})=\frac{\frac{R_k}{H_k}}{C(\boldsymbol{R})}\quad\quad k=1,\ldots,N
\end{cases} 
\label{Stationary}
\end{equation}
and $C(\boldsymbol{R})$ is a normalization factor that is equal to:
\begin{equation}
C(\boldsymbol{R})=1+\sum_{k=1}^{N}\frac{R_k}{H_k}
\label{normalizationfactor}
\end{equation}
\label{stationarydist}
\end{proposition}
\vspace{-17pt}
\begin{IEEEproof} It is sufficient to show that the preceding distribution verifies the detailed balance equations \cite{norris_1997}. Consider the two states $0$ and $k$: we have from (\ref{Stationary}) that $\frac{\overline{\pi}_{k}}{\overline{\pi}_{0}}=\frac{R_k}{H_k}$ which is exactly the detailed balance equation between states $0$ and $k$. This holds for any of the states $k$ and therefore the proof is complete.
\end{IEEEproof}
\begin{theorem}
In the aforementioned system, the average age of the system is:
\begin{equation}
\overline{\Delta}(\boldsymbol{R})=\sum_{k=1}^{N}\frac{1}{\lambda_k}-\sum_{k=1}^{N}\frac{1}{H_k}+N\frac{\sum\limits_{k=1}^{N}\frac{R_k}{H^2_k}}{C(\boldsymbol{R})}+C(\boldsymbol{R})(\sum_{k=1}^{N}\frac{1}{R_k})
\end{equation}
\end{theorem}
\begin{IEEEproof}
By considering the stationary distribution reported in Proposition \ref{stationarydist} and both the transitions of Table \ref{trans} and the differential equation vector of (\ref{bstuff}), we can apply Theorem \ref{theomhem} to find the vector $\overline{\boldsymbol{v}}_{q}=[\overline{v}_{q0},\overline{v}_{q1}]\:\: \forall q\in\mathbb{Q}$. By doing so, we end up with the following set of equations:
\begin{equation}
\overline{v}_{00}(\sum_{k=1}^{N}R_k)=\overline{\pi}_0+ \sum_{\substack{k=1 \\ k\neq i}}^{N}H_k\overline{v}_{k0}+H_i\overline{v}_{i1}
\label{00}
\end{equation}
\begin{equation}
\overline{v}_{01}(\lambda_i+\sum_{k=1}^{N}R_k)=\overline{\pi}_0+ \sum_{\substack{k=1 \\ k\neq i}}^{N}H_k\overline{v}_{k1}+H_i\overline{v}_{i1}
\label{01}
\end{equation}
\begin{equation}
\overline{v}_{k0}(H_k)=\overline{\pi}_k+ R_k\overline{v}_{00} \quad\quad k=1,\ldots,N
\label{k0}
\end{equation}
\begin{equation}
\overline{v}_{k1}(\lambda_i+H_k)=\overline{\pi}_k+ R_k\overline{v}_{01} \quad\quad k=1,\ldots,N
\label{k1}
\end{equation}
As previously stated in Theorem \ref{theomhem}, the average age of the link of interest $i$ is $\overline{\Delta}_i(\boldsymbol{R})=\sum\limits_{k=0}^{N}\overline{v}_{k0}$. The first step to solve this set of $2N+2$ equations consists of calculating $\overline{v}_{01}$. From the $N$ equations presented in (\ref{k1}), we can conclude that $\overline{v}_{k1}=\frac{\overline{\pi}_k}{\lambda_i+H_k}+\frac{R_k}{\lambda_i+H_k} \quad \forall k=1,\ldots,N$. By replacing these values in (\ref{01}), we can conclude that:
\begin{equation}
\overline{v}_{01}=\frac{\overline{\pi}_0+\sum\limits_{k=1}^{N}\frac{H_k\overline{\pi}_k}{\lambda_i+H_k}}{\lambda_i+\sum\limits_{k=1}^{N}R_k(1-\frac{H_k}{\lambda_i+H_k})}
\end{equation}
Knowing that $\overline{v}_{i1}=\frac{\overline{\pi}_i}{\lambda_i+H_i}+\frac{R_i\overline{v}_{01}}{\lambda_i+H_i}$, we can therefore proceed to compute $\overline{v}_{00}$ using (\ref{00}):
\begin{equation}
\overline{v}_{00}=\frac{1-\overline{\pi}_i+\frac{H_i\overline{\pi}_i}{\lambda_i+H_i}}{R_i}+\frac{(\frac{H_i}{\lambda_i+H_i})(\overline{\pi}_0+\sum\limits_{k=1}^{N}\frac{H_k\overline{\pi}_k}{\lambda_i+H_k})}{\lambda_i+\sum\limits_{k=1}^{N}R_k(1-\frac{H_k}{\lambda_i+H_k})}
\end{equation}
The next step consists of using the results of (\ref{Stationary}). By replacing $\overline{\pi}_k\:\: 0\leq k\leq N$ by their values and after algebraic manipulations, we get that:
\begin{equation}
\overline{v}_{00}=\frac{1}{R_i}+\frac{1}{C(\boldsymbol{R})}(\frac{1}{\lambda_i}-\frac{1}{H_i})
\end{equation}
Next, we know that $\overline{\Delta}_i(\boldsymbol{R})=\overline{v}_{00}+\sum\limits_{k=1}^{N}\overline{v}_{k0}$. By using the $N$ equations of (\ref{k0}), we get that $\sum\limits_{k=1}^{N}\overline{v}_{k0}=\sum\limits_{k=1}^{N}\frac{\overline{\pi}_k}{H_k}+\overline{v}_{00}(C(\boldsymbol{R})-1)$ and the average age of the link of interest $i$ is therefore: 
\begin{equation}
\overline{\Delta}_i(\boldsymbol{R})=\frac{C(\boldsymbol{R})}{R_i}+\frac{1}{\lambda_i}-\frac{1}{H_i}+\frac{\sum\limits_{k=1}^{N}\frac{R_k}{H^2_{k}}}{C(\boldsymbol{R})}
\end{equation}
These results are general and hold for any link $i$ in the network. Therefore, by summing over all the links $i$, the total average age of the network can be concluded.
\end{IEEEproof}
\subsection{Contention Resolution}
In realistic scenarios, the sensing delay cannot be neglected and therefore the back-off 
time is chosen as a multiple 
of mini-slots where the duration of the mini-slot $T_{slot}$ is dictated by physical limitations such as propagation delay (the time necessary for the receiver to detect the radio signals). In fact, link $k$ 
picks a random back-off time uniformly distributed from the range $[0,W_k-1]$ where $W_k$ is referred to as the Contention Window (\textbf{CW}) of link $k$. In this case, the average back-off time becomes $T_{slot}\frac{W_k-1}{2}$. Knowing that our proposed analysis assumes an average back-off time of $\frac{1}{R_k}$, we therefore choose a contention window such that:
\begin{equation}
W_k=\frac{2}{T_{slot}R_k}+1
\label{contentionwindow}
\end{equation}
In order to have a reasonably low collision probabilities, we proceed to lowerbound the contention window of each link by a sufficiently high enough minimal contention window $W_0$ (i.e. $W_k\geq W_0\; \forall k$). By doing so, collisions can be ignored and the performance of the scheme coincides with the idealized CSMA settings. This can be thought to be similar to the well-known Binary Exponential Back-off (\textbf{BEB}) mechanism in IEEE 802.11 where the contention window is increased to avoid collisions in the network. However in our approach, unlike the latter, a successful transmission will not reset the contention window to its initial value but rather keep the same \emph{sufficiently high} CW value in the aim of avoiding collisions between links. Consequently, to include the contention resolution aspect of the CSMA in our analysis, it is enough to upperbound the back-off rate $R_k$ of each link by:
\begin{equation}
R_{UB}=\frac{2}{(W_0-1)T_{slot}}
\end{equation}
\color{black}
\subsection{Average age minimization}
The ultimate goal of the work is to optimize the average back-off time of each link in a way to minimize the total average age of the network. We can therefore formulate our optimization problem as follows:
\begin{equation}
\begin{aligned}
& \underset{\boldsymbol{R}}{\text{minimize}}
&& \sum_{k=1}^{N}\frac{1}{\lambda_k}-\sum_{k=1}^{N}\frac{1}{H_k}+N\frac{\sum\limits_{k=1}^{N}\frac{R_k}{H^2_k}}{C(\boldsymbol{R})}+C(\boldsymbol{R})(\sum_{k=1}^{N}\frac{1}{R_k}) \\
& \text{subject to}
& & 0\leq R_k\leq R_{UB} \quad\quad k = 1, \ldots, N
\end{aligned}
\label{convexop}
\end{equation}
\begin{remark}
An important thing we notice is the fact that the total average age depends on the arrival rate of each link solely through the term $\sum_{k=1}^{N}\frac{1}{\lambda_k}$. This means that the optimization of the average back-off timers is \emph{independent} of the arrival rate which makes it interesting to be implemented in practice as there is no need to have any prior knowledge on the arrival rate of each link to calibrate their average back-off time.
\end{remark}
At a first glimpse, the problem in (\ref{convexop}) appears to be a special case of the well-known sum of ratio problems which are generally hard to solve \cite{doi:10.1080/1055678031000105242}. However, by analyzing its structure, this optimization problem can be converted into an equivalent convex problem via variable substitutions as will be depicted in the following. To put this into perspective, let us introduce the new variable $\epsilon=\frac{1}{C(\boldsymbol{R})}$ and the variables $f_k$ such that $f_k=\epsilon R_k \:\:\forall k$. The optimization problem becomes:
\begin{equation}
\begin{aligned}
& \underset{\boldsymbol{f},\epsilon}{\text{minimize}}
&& \sum_{k=1}^{N}\frac{1}{\lambda_k}-\sum_{k=1}^{N}\frac{1}{H_k}+N\sum_{k=1}^{N}\frac{f_k}{H^2_k}+\sum_{k=1}^{N}\frac{1}{f_k}\\
& \text{subject to}
& & 0\leq f_k\leq \epsilon R_{UB} \quad\quad k = 1, \ldots, N \\
& &  & \frac{1}{1+\sum\limits_{k=1}^{N}\frac{R_{UB}}{H_k}}\leq\epsilon\leq1\\
& &  &  \sum_{k=1}^{N}\frac{f_k}{H_k}=1-\epsilon
\end{aligned}
\label{convexopredefined}
\end{equation}
The formulated problem is convex in $\boldsymbol{f}=[f_1,\ldots,f_N]$ and $\epsilon$ as the objective function is the sum of convex functions in $(\boldsymbol{f},\epsilon)$ and the constraints are linear. We therefore formulate the Lagrange function as follows:
\begin{multline}
\mathcal{L(\boldsymbol{f},\epsilon,\nu,\gamma,\rho,\boldsymbol{\mu},\boldsymbol{\eta})}=\sum_{k=1}^{N}\frac{1}{\lambda_k}-\sum_{k=1}^{N}\frac{1}{H_k}+N\sum_{k=1}^{N}\frac{f_k}{H^2_k}\\+\sum_{k=1}^{N}\frac{1}{f_k}+\nu(\epsilon-1)+\gamma(\frac{1}{1+\sum\limits_{k=1}^{N}\frac{R_{UB}}{H_k}}-\epsilon)\\+\sum_{k=1}^{N}\mu_k(f_k-\epsilon R_{UB})+\sum_{k=1}^{N}\eta_k(-f_k)+\rho(\sum_{k=1}^{N}\frac{f_k}{H_k}+\epsilon-1)
\label{lagrangefunction}
\end{multline}
As the problem was shown to be convex, we formulate the Karush$–-$Kuhn$–-$Tucker (\textbf{KKT}) conditions which are sufficient for optimality in our case:
\begin{equation}
\frac{N}{H_k^2}-\frac{1}{f_k^{*^2}}+\mu_k^*-\eta_k^*+\frac{\rho^*}{H_k}=0 \quad\quad k=1,\ldots,N
\label{KKT1}
\end{equation}
\vspace{-15pt}
\begin{equation}
\rho^*+\nu^*-\gamma^*-R_{UB}\sum\limits_{k=1}^{N}\mu_k^*=0
\label{KKT2}
\end{equation}
\vspace{-10pt}
\begin{equation}
\gamma^*(\frac{1}{1+\sum\limits_{k=1}^{N}\frac{R_{UB}}{H_k}}-\epsilon^*)=0
\label{KKT3}
\end{equation}
\begin{equation}
\nu^{*}(\epsilon^{*}-1)=0
\label{KKT4}
\end{equation}
\vspace{-15pt}
\begin{equation}
\mu_k^*(f_k^*-\epsilon^* R_{UB})=0 \quad\quad k=1,\ldots,N
\label{KKT5}
\end{equation}
\vspace{-15pt}
\begin{equation}
\eta_k^{*}f^*_k=0 \quad\quad k=1,\ldots,N
\label{KKT6}
\end{equation}
\vspace{-15pt}
\begin{equation}
\nu^*,\gamma^*\geq0\quad\quad \boldsymbol{\mu},\boldsymbol{\eta}\geq\boldsymbol{0}
\label{KKT7}
\end{equation}
\vspace{-15pt}
\begin{equation}
0\leq f^*_k\leq \epsilon^* R_{UB} \quad\quad k = 1, \ldots, N
\label{KKT8}
\end{equation}
\vspace{-15pt}
\begin{equation}
\frac{1}{1+\sum\limits_{k=1}^{N}\frac{R_{UB}}{H_k}}\leq\epsilon^*\leq1
\label{KKT9}
\end{equation}
\begin{equation}
\sum_{k=1}^{N}\frac{f^*_k}{H_k}=1-\epsilon^*
\label{KKT10}
\end{equation}
In the following, we find the optimal solution using the above \emph{sufficient} optimality conditions. First we suppose that $\nu^*>0$, which means that $\epsilon^*=1$. Replacing this in eq. (\ref{KKT10}), and knowing that $f_k^*\geq0\:\forall k$, we get that $f_k^*=0\:\forall k$ and the objective function tends to $+\infty$ which is surely not optimal. Therefore, we can conclude that $\nu^*=0$. The same argument can be used to show that $\eta_k=0\:\forall k$. We now suppose that $\gamma^*>0$, which entails that $\epsilon^*=\frac{1}{1+\sum\limits_{k=1}^{N}\frac{R_{UB}}{H_k}}$. By replacing this in eq. (\ref{KKT10}), and by noting the conditions in eq. (\ref{KKT8}), we can conclude that this is only feasible when $f_k^*=\epsilon^*R_{UB}\:\forall k$. We replace $f_k^*$ and $\epsilon^*$ by their values in eq. (\ref{KKT1}) and we end up with:
\begin{equation}
\mu_k^*=\frac{-\rho^*}{H_k}+\frac{\Big(1+\sum\limits_{k=1}^{N}\frac{R_{UB}}{H_k}\Big)^2}{R_{UB}^2}+\frac{-N}{H_k^2} \quad k=1,\ldots,N
\end{equation}
Using the fact that $\mu_k^*\geq0$, we have the following $N$ conditions on $\rho^*$:
\begin{equation}
\rho^*\leq \frac{H_k\Big(1+\sum\limits_{k=1}^{N}\frac{R_{UB}}{H_k}\Big)^2}{R^2_{UB}}-\frac{N}{H_k}\quad k=1,\ldots,N
\label{condrho1}
\end{equation} 
Knowing that $\mu_k\geq0$, and by using eq. (\ref{KKT2}), we can conclude that $\rho^*\geq\gamma^*>0$. To proceed with this case, we define $x=\sum\limits_{k=1}^{N}\frac{1}{H_k}$ and $y=\sum\limits_{k=1}^{N}\frac{1}{H^2_k}$. By summing eqs. (\ref{KKT1}) for all $k$ and using the results of eq. (\ref{KKT2}), we end up with:
\begin{equation}
\gamma^*=NyR_{UB}-\frac{N\Big(1+\sum\limits_{k=1}^{N}\frac{R_{UB}}{H_k}\Big)^2}{R_{UB}}+\rho^*(1+xR_{UB})
\end{equation}
As $\gamma^*>0$, the final condition on $\rho^*$ is therefore:
\begin{equation}
\rho^*>\frac{N(1+xR_{UB})}{R_{UB}}-\frac{NyR_{UB}}{1+xR_{UB}}
\label{condrho2}
\end{equation}
If $\exists\rho^*$ such that the conditions reported in eqs. (\ref{condrho1}) and (\ref{condrho2}) are verified then $f_k^*=\epsilon^*R_{UB}\:\forall k$ and the original problem's optimal point is: $R_k=R_{UB}\:\forall k$. We will show that this is always achieved when $H_k=H \:\forall k$ in the next Lemma. In the latter cases where $\gamma^*=0$, this entails that there could be at least one link $k$ such that $f^*_k<\epsilon^*R_{UB}$. Therefore, in this scenario, the optimal solution $f_k^*\:\: \forall k$ is such that:
\begin{equation}
f_k^*=\begin{cases}
\epsilon^*R_{UB} & \text{if}\: \mu_k^*>0 \\
\sqrt{\frac{H_k}{\frac{N}{H_k}+\rho^*}} & \text{if} \: \mu_k^*=0 
\end{cases}
\end{equation}
where $\rho^*$ and $\epsilon^*$ verify: 
\begin{equation*}
\rho^*=R_{UB}\sum\limits_{k=1}^{N}\mu_k^*
\end{equation*}
\begin{equation}
\epsilon^*=1-\sum_{k=1}^{N}\frac{f^*_k}{H_k}
\end{equation}
The optimal solution $(\boldsymbol{f}^*,\epsilon^*)$ of the problem in (\ref{convexopredefined}) can therefore be found. The optimal back-off rate of each link $k$ can then be deduced $R^*_k=\frac{f^*_k}{\epsilon^*}\:\forall k$. 
\begin{lemma}
If $H_k=H\:\:\forall k$, then $R^*_k=R_{UB}\:\:\forall k$
\label{samechannel}
\end{lemma}
\begin{IEEEproof}
The proof can be found in Appendix A.
\end{IEEEproof}
\begin{remark}
The results of Lemma \ref{samechannel} are interesting as they highlight the fact that in the case where links have the same average channel holding time $\frac{1}{H}$, no priority is given to any of the links in accessing the channel. In other words, when links have similar average channel conditions, power and packet length, then there is no need to optimize the average back-off duration of each link. In fact, it is sufficient to make all links be as aggressive on the channel as possible while making sure collisions are handled (i.e. $R_k=R_{UB}\:\forall k$). On the other hand, in the asymmetric average channel holding time scenario, links with a smaller average channel holding time will be given priority to access the channel by assigning to them a smaller average back-off time as will be highlighted in the next section.
\end{remark}
\section{Numerical Results}
As pointed out in our theoretical analysis, the optimal back-off rate $\boldsymbol{R}^*$ is independent of the arrival rate of each link. We therefore fix the arrival rate to $\lambda_k=1\:\:\forall k$ in the sequel. 

\begin{figure}[!ht]
\centering
\includegraphics[width=.99\linewidth]{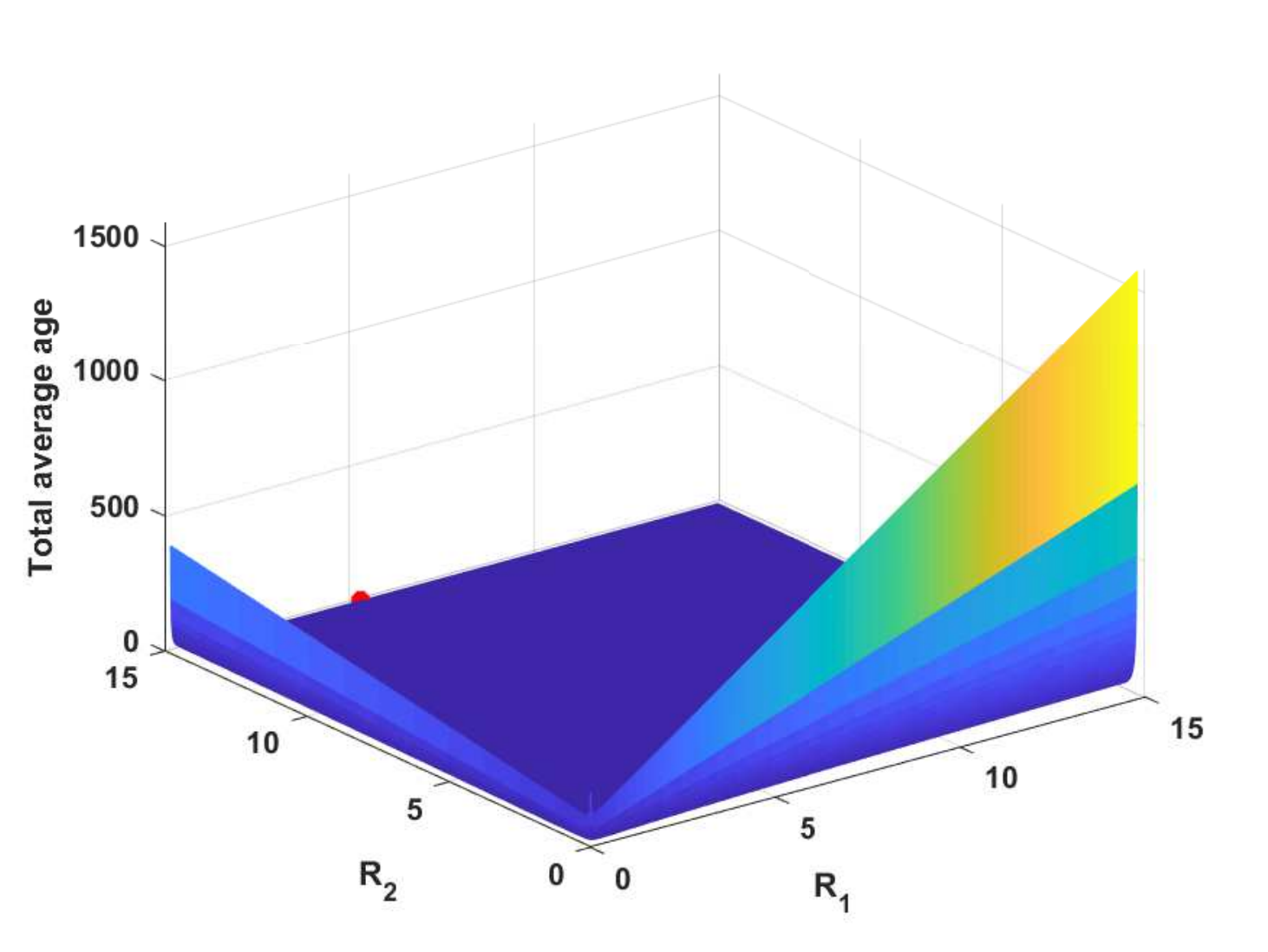}
\caption{Illustration of the total average age of the system in function of $R_1$ and $R_2$}
\setlength{\belowcaptionskip}{-5pt}
\label{2d}
\end{figure}

In the first scenario, we consider $2$ links with link $1$ and $2$ having an average channel holding time of $1$ ms and $0.2$ ms respectively. We have set the contention window lowerbound to $W_0=16$. The slot time is set to $T_{s}=9 \mu s$ (as adopted in IEEE 802.11n \cite{5307322}) which leads to $R_{UB}=14.8$. We report the total average age of the system in function of $R_1$ and $R_2$ in Fig. \ref{2d}. The first thing we can see is that if both $R_1$ and $R_2$ approach zero, the total average age is high as links barely access the channel. Also, if only one of them approaches zero, the total average age grows rapidly due to starvation of that link. The optimal total average age of the network (marked in red) was achieved for $(\overline{\Delta}^*=4.44,R^*_1=5.16,R^*_2=14.8)$. This gives us an interesting conclusion: links with a small channel holding time (faster transmission) should be given a higher priority to access the channel by increasing their aggressiveness on the channel (smaller average back-off time).

In the second scenario, we consider a realistic implementation of our proposed algorithm in an IEEE 802.11 environment with a slot time of $T_{s}=9 \mu s$. We consider an access point communicating to $N$ interfering nodes. Each link in the network has an average channel holding time of $1$ ms. The access point, with knowledge of the average service time of each node, solves the convex problem in (\ref{convexopredefined}) and disseminate for each node its optimal back-off window. We vary the density of the nodes and adjust the contention window's lowerbound $W_0$ accordingly. 
 We use CVX \cite{cvx} to solve the unconstrained idealized CSMA settings with no collisions and compare it with our IEEE 802.11 implementation results. We can see that the performance of our IEEE 802.11 implementation virtually coincide with the unconstrained idealized CSMA settings for low density cases. As the density of the nodes increases, a gap starts forming between the realistic IEEE 802.11 implementation and the idealized CSMA settings. This is due to the fact that as nodes' density increases, the effect of collisions will have a more severe impact on the performance. However, one can see that even for a high density environment of $8$ nodes in a single collision domain, the performance degradation with respect to the perfect idealized CSMA settings is only around $10\%$. 
\begin{figure}[!ht]
\centering
\includegraphics[width=.9\linewidth]{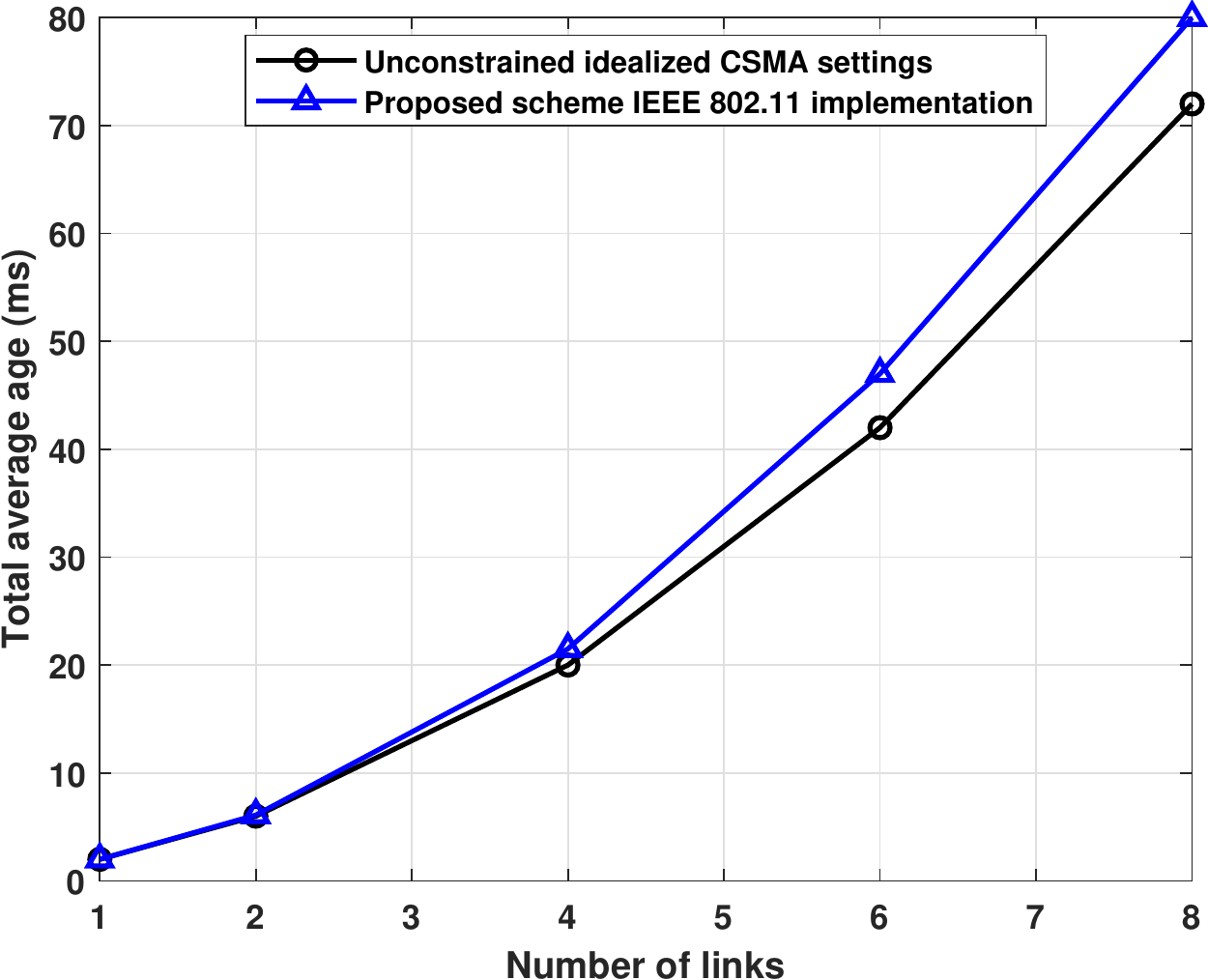}
\caption{The optimal total average age of the system in function of the nodes' density}
\label{density}
\end{figure}

In the third scenario, we aim to compare the optimized CSMA scheme to a simple Round Robin scheme (\textbf{RR}). As seen in Table \ref{RRCSMA}, the optimized CSMA outperforms the RR scheme when $(H_1=0.1,H_2=10)$. However, although the CSMA scheme has been optimized in the aim of minimizing the average AoI of the network, it lacks behind the RR scheme in the case of symmetrical service rate. Unlike the works in the throughput maximization literature where typical CSMA schemes were shown to achieve the maximum achievable throughput of the network when optimized \cite{6406138}, this is not true for the case of minimizing the AoI. This suggests that for optimizing the average AoI, new distributed schemes (e.g. modified CSMA schemes) have to be considered to achieve better age performance than the standard CSMA scheme.
\def\arraystretch{1.5}
\begin{center}
\begin{tabular}{|c|c|c|}

 \hline
 Service Rate & $\overline{\Delta}_{CSMA}$ & $\overline{\Delta}_{RR}$ \\
  \hline
 $(H_1=1,H_2=1)$  & $6$ & $5$ \\
 $(H_1=0.1,H_2=10)$  & $20.5$ & $22.3$\\
 \hline
\end{tabular}
 \captionof{table}{Comparison between CSMA and Round Robin}
 \label{RRCSMA}
\end{center}
\color{black}
\section{Conclusion}
In this paper, we have investigated the AoI in a CSMA environment where $N$ links contend for the channel. Using SHS tools, we were able to find a closed form of the total average age of the network. An optimization problem with the aim of minimizing the total average age of the network was then formulated. By investigating it, an equivalent convex formulation was presented that makes finding the optimal back-off time of each link a simple task. Numerical implementations of our proposed solution in an IEEE 802.11 network was provided and its performance was highlighted in function of the nodes' density. We have also shown that, although optimized, the standard CSMA scheme still lacks behind other distributed schemes in terms of average age in some special cases. These results suggest the necessity to find new distributed schemes to further minimize the average age of any general network.
\bibliographystyle{IEEEtran}
\bibliography{trialout}
\appendices
\section{Proof of Lemma 1}
To prove this, it is sufficient to show that the conditions on $\rho^*$ of eqs. (\ref{condrho1}) and (\ref{condrho2}) are always verified for all feasible $(N,H,R_{UB})$. More specifically, we have to show that:
\begin{equation}
\frac{H\Big(1+\frac{NR_{UB}}{H}\Big)^2}{R^2_{UB}}-\frac{N}{H}>\frac{N(1+\frac{NR_{UB}}{H})}{R_{UB}}-\frac{\frac{N^2R_{UB}}{H^2}}{1+\frac{NR_{UB}}{H}}
\end{equation}
By taking a common denominator, and knowing that $H,R_{UB}>0$, the positivity of the aforementioned expression is equivalent to that of the following cubic polynomial $f(N,H,R_{UB})$:
\begin{equation}
(H+NR_{UB})^3-NR(H+NR_{UB})^2-NHR_{UB}^2>0
\end{equation} 
By deriving $f(N,H,R_{UB})$ with respect to $H$, we find that:
\begin{equation}
\frac{\partial f(N,H,R_{UB})}{\partial H}=3H^2+H(4NR_{UB})+N^2R_{UB}^2-NR_{UB}^2
\end{equation}
To study the sign of $\frac{\partial f(N,H,R_{UB})}{\partial H}$, we derive one more time with respect to $H$ and we can show that:
\begin{equation}
\frac{\partial^2 f(N,H,R_{UB})}{\partial H^2}=6H+4NR_{UB}>0
\end{equation}
Therefore, $\frac{\partial f(N,H,R_{UB})}{\partial H}$ is always increasing, baring in mind that:
\begin{equation}
\lim_{H\to 0} \frac{\partial f(N,H,R_{UB})}{\partial H}=N^2R_{UB}^2-NR_{UB}^2\geq0
\end{equation}
since $N\geq1$. Hence, we can conclude that $f(N,H,R_{UB})$ increases with $H$. By following the same approach, we can find that: $\frac{\partial f(N,H,R_{UB})}{\partial R_{UB}}=2NH^2+2N^2HR_{UB}-2NHR_{UB}$ and $\frac{\partial^2 f(N,H,R_{UB})}{\partial R_{UB}^2}=2NH(N-1)\geq0$ since $N\geq1$. As $\lim_{R_{UB}\to 0}\frac{\partial f(N,H,R_{UB})}{\partial R_{UB}}=2NH^2>0$, we can conclude that $f(N,H,R_{UB})$ increases with $R_{UB}$. We can use the same argument over $N$ by relaxing its discrete nature to a continuous one $n\in [1,+\infty]$ and noting that: $\frac{\partial f(n,H,R_{UB})}{\partial n}=2R_{UB}H^2+2nR_{UB}^2H-R_{UB}^2H$, $\frac{\partial^2 f(n,H,R_{UB})}{\partial n^2}=2R_{UB}^2H$ and $\frac{\partial f(n,H,R_{UB})}{\partial n}\Big|_{n=1}=2R_{UB}H^2+R_{UB}^2H$, we can show that $f(N,H,R_{UB})$ increases with $N$. Knowing that $f(1,H,R_{UB})$ increases with $(H,R_{UB})$ and that $\lim_{(R_{UB},H)\to (0,0)}f(1,H,R_{UB})\longrightarrow0$, we can therefore assess that $\forall\delta_1,\delta_2>0$ such that $H\geq\delta_1,R_{UB}\geq\delta_2$, $\exists\delta>0$ such that $f(N,H,R_{UB})\geq\delta$ with $\delta=(\delta_1+\delta_2)^3-\delta_2(\delta_1+\delta_2)^2-\delta_2^2(\delta_1+\delta_2)+\delta_2^3>0$ which concludes our proof.

\end{document}